\documentclass[aps,prl,twocolumn,groupedaddress,showpacs,superscriptaddress]{revtex4}

\usepackage{graphicx}
\usepackage{epsfig}
\usepackage{dcolumn}
\usepackage{amsmath}
\usepackage{graphicx}
\usepackage{amssymb}
\usepackage{bm}
\usepackage{array}
\usepackage{hyperref}
\usepackage{slashed}
\usepackage{braket}

\begin{document}

\title{The $\cos2\phi$ azimuthal asymmetry of unpolarized $p\bar{p}$ collisions at Tevatron}

\newcommand*{\PKU}{School of Physics and State Key Laboratory of Nuclear Physics and
Technology, Peking University, Beijing 100871,
China}\affiliation{\PKU}
\newcommand*{\CHEP}{Center for High Energy
Physics, Peking University, Beijing 100871,
China}\affiliation{\CHEP}

\author{Tianbo Liu}\affiliation{\PKU}
\author{Bo-Qiang Ma}\email{mabq@pku.edu.cn}\affiliation{\PKU}\affiliation{\CHEP}

\date{\today}

\begin{abstract}
We calculate the $\cos2\phi$ azimuthal asymmetry of the unpolarized
$p\bar{p}$ Drell-Yan dilepton production process in the
$Z$-resonance region at the Tevatron kinematic domain. Such an
azimuthal asymmetry can provide additional information about a
spin-related new parton distribution function, i.e., the
Boer-Mulders function of the proton, compared to the $pp$ process.
Therefore the available data of unpolarized proton-antiproton
collision at Tevatron can contribute to our study on the spin
structure of the nucleon.
\end{abstract}


\maketitle

The study of the intrinsic transverse momentum dependent (TMD)
distribution functions has received much attention in recent
years~\cite{Barone:2001sp}. Such new quantities of the nucleon
provide us a significant perspective on understanding the spin
structure of hadrons and the non-perturbative properties of quantum
chromodynamics (QCD). The intrinsic transverse momentum of partons
may cause special effects in high energy scattering
experiments~\cite{Cahn:1978se}. It was naively speculated that the
polarization of at least one incoming hadron is necessary to
investigate the spin-related structure and properties of hadrons,
however the situation will change if the transversal motions of
quarks inside the hadron will take into account. The Drell-Yan
process is an ideal ground for testing perturbative QCD and probing
TMD distribution functions, and its cross section is well described
by next-to-leading order QCD calculations~\cite{Stirling:1993gc}.
Surprisingly, the first measurement of the Drell-Yan angular
distribution, performed by NA10 Collaboration for $\pi N$ at 140,
194 and 286~GeV, indicates a sizable $\cos2\phi$ azimuthal
asymmetry~\cite{Falciano:1986wk,Guanziroli:1987rp} which cannot be
described by leading and next-to-leading order perturbative
QCD~\cite{Brandenburg:1993cj}. Furthermore, the subsequent result by
the Fermilab E615 Collaboration reveals that the Lam-Tung
relation~\cite{Lam:1978pu}, which is analogous to the Callan-Gross
relation~\cite{Callan:1969uq} in deep-inelastic scattering, obtained
as a consequence of the spin-$\frac{1}{2}$ nature of the quarks, is
clearly violated~\cite{Conway:1989fs}. The violation has also been
tested in recent $pd$ and $pp$ Drell-Yan dimuon processes measured
by E866/NuSea Collaboration~\cite{Zhu:2006gx,Zhu:2008sj}.

Several attempts were made to interpret this asymmetry, such as the
factorization breaking QCD vacuum effect~\cite{Brandenburg:1993cj}
which is possible the helicity flip in the instanton
model~\cite{Boer:2004mv}, higher twist
effect~\cite{Brandenburg:1994wf,Eskola:1994py,Heinrich:1991zm} and
the coherent states~\cite{Blazek:1989kt}. Boer pointed out that the
$\cos2\phi$ azimuthal asymmetry could be due to a non-vanished TMD
distribution function $h_1^\perp(x,\bm{p}_T^2)$~\cite{Boer:1999mm},
named as the Boer-Mulders function later, as one of the eight
leading-twist TMD distribution function contained
in~\cite{Mulders:1995dh,Boer:1997nt}
\begin{equation}
\begin{split}
\Phi = & \frac{1}{2} \bigg\{ f_1\slashed{n}_+ - f_{1T}^\perp \frac{\epsilon_T^{ij} p_{Ti} S_{Tj}}{M}\slashed{n}_+  + h_{1T}\frac{[\slashed{S}_T,\slashed{n}_+]\gamma_5}{2} \\
& + \Big(S_L g_{1L} + \frac{\bm{p}_T \cdot \bm{S}_T}{M} g_{1T}\Big) \gamma_5\slashed{n}_+ \\
&  + \Big(S_L h_{1L}^\perp + \frac{\bm{p}_T \cdot \bm{S}_T}{M} h_{1T}^\perp\Big) \frac{[\slashed{p}_T,\slashed{n}_+]\gamma_5}{2M} \\
& + i h_1^\perp \frac{[\slashed{p}_T,\slashed{n}_+]}{2M} \bigg\},
\end{split}
\end{equation}
where $\Phi$ is the quark-quark correlation matrix, defined as
\begin{equation}
\Phi_{ij}(p, P, S) = \int \frac{d^4\xi}{(2\pi)^4} e^{ip\cdot \xi} \braket{P,S | \bar{\psi}_j(0) \mathcal{W}[0,\xi]\psi_i(\xi) | PS}.
\end{equation}
The Boer-Mulders function is another time-reversal odd ($T$-odd)
distribution function which characterizes the correlation between
quark transverse momentum and quark transverse spin, analogous to
the Sivers function $f_{1T}^\perp(x,\bm{p}_T^2)$ which signifies the
correlation between quark transverse momentum and hadron transverse
spin~\cite{Sivers:1989cc}. The non-vanished $T$-odd distribution
functions can arise from the initial-state or final-state
interaction~\cite{Collins:2002kn,Brodsky:2002rv,Gamberg:2003ey,Boer:2002ju}.
In general, the path-order Wilson line arising from the requirement
of QCD gauge invariance for quark correlation functions provides
non-trivial phases and leads to non-vanished $T$-odd distribution
functions~\cite{Ellis:1982wd,Efremov:1979qk,Collins:1981uw,Ji:2002aa}.
Due to the present of Wilson line, opposite sign of the Boer-Mulders
function or Sivers function in semi-inclusive deep inelastic
scattering (SIDIS) and Drell-Yan processes is
expected~\cite{Boer:2003cm,Collins:2004nx},
\begin{equation}
h_1^\perp(x, \bm{p}_T^2)|_{\mathrm{SIDIS}} = - h_1^\perp(x, \bm{p}_T^2)|_{\mathrm{DY}}.
\end{equation}
The existence of $T$-odd distribution function can cause azimuthal
asymmetries in SIDIS at leading twist level~\cite{Boer:1997nt}, and
the product of two Boer-Mulders functions of two incoming hadrons
may give a sizable $\cos2\phi$ azimuthal asymmetry in unpolarized
Drell-Yan processes by establishing a preferred transverse momentum
direction from the spin-transverse momentum correlation, which is
called the Boer-Mulders effect~\cite{Boer:1999mm}. Thus, the
measurement of the Boer-Mulders function will promote our
understanding of QCD. Many theoretical and phenomenological studies
are carried out along this
direction~\cite{Lu:2004hu,Lu:2005rq,Bianconi:2005bd,Sissakian:2005vd,
Sissakian:2005yp,Lu:2006ew,Barone:2006ws,Lu:2007kj,Gamberg:2005ip,Zhang:2008nu,
Zhang:2008ez,Barone:2010gk,Lu:2009ip,Lu:2011mz,Yuan:2003wk,Pasquini:2006iv,Gockeler:2006zu,Burkardt:2007xm}.

Recently, the Collider Detector at Fermilab (CDF) Collaboration
first measured the angular distribution coefficients of Drell-Yan
$e^+e^-$ pairs in the $Z$ mass region from unpolarized $p\bar{p}$
collisions $p+\bar{p}\rightarrow\gamma^*/Z+X\rightarrow l^+l^-+X$ at
$\sqrt{s}=1.96~\textrm{TeV}$~\cite{Aaltonen:2011nr}. This indicates
that it is feasible to investigate spin physics at Tevatron. In this
paper, we calculate the $\cos2\phi$ azimuthal asymmetry caused by
the Boer-Mulders effect in the $Z$-pole region with the kinematic
conditions at Tevatron.

The angular distribution coefficients are generally frame dependent.
We choose the Collins-Soper (CS) frame~\cite{Collins:1977iv} to
perform the calculation. It is the center of mass of the lepton pair
with the $z$ axis defined as the bisector of $p$ and $\bar{p}$
beams. The polar angular $\theta$ is defined as the angular of the
positive lepton with respect to the $z$ axis direction, and the
azimuthal angular $\phi$ is defined as the angular of the lepton
plane with respect to the proton plane. In this frame the Lam-Tung
relation is insensitive to the higher fixed-order perturbative
QCD~\cite{Mirkes:1994eb} or the QCD
resummation~\cite{Boer:2006eq,Berger:2007si,Berger:2007jw}. The
angular differential cross section for unpolarized Drell-Yan process
has the general form:
\begin{align}\label{diffcrosssection}
\frac{1}{\sigma}\frac{d\sigma}{d\Omega}=&\frac{3}{4\pi}\left.\frac{1}{\lambda+3}\right(1+\lambda\cos^2\theta
+\mu\sin2\theta\cos\phi\nonumber\\
&+\left.\frac{\nu}{2}\sin^2\theta\cos2\phi\right),
\end{align}
where $\Omega$ is the solid angle and $\lambda$, $\mu$, and $\nu$
are angular distribution coefficients. For azimuthal symmetrical
scattering, the coefficients $\mu=\nu=0$. It can also be written
as~\cite{Oakes:NCA44,Lam:1978pu}:
\begin{align}\label{diffcrosssectionW}
\frac{d\sigma}{d\Omega}=&W_T(1+\cos^2\theta)+W_L(1-\cos^2\theta)\nonumber\\
&+W_\Delta\sin2\theta\cos\phi+W_{\Delta\Delta}\sin^2\theta\cos2\phi.
\end{align}

When taking into account both virtual photon and $Z$-boson
contribution, the leading order unpolarized Drell-Yan cross section
is~\cite{Boer:1999mm}
\begin{equation}\label{diffBoer}
\begin{split}
\frac{d\sigma(h_1h_2\rightarrow l\bar{l}X)}{d\Omega dx_1dx_2d^2\bm{q}_T}&=\frac{\alpha^2}{3Q^2}
\sum_a\left\{K_1(\theta)\mathcal{F}[f_{1a}f_{1a}]\frac{}{}\right.\\
&+[K_3(\theta)\cos2\phi+K_4(\theta)\sin2\phi]\\
\times&\left.\mathcal{F}
\left[(2\hat{\bm{h}}\cdot\bm{p}_T\hat{\bm{h}}\cdot\bm{k}_T-\bm{p}_T\cdot\bm{k}_T)
\frac{h_{1a}^{\perp}h_{1a}^{\perp}}{M^2}\right]\right\},
\end{split}
\end{equation}
where $x_1$, $x_2$ are the Bjorken variables standing for the longitudinal
momentum fractions carried by the partons in the proton and antiproton, and
$\alpha$, $M$, $\bm{q}_T$, and $Q$ are the fine structure constant, the mass of proton,
the transverse momentum, and invariant mass of $\gamma^*/Z$ respectively.
The structure function notation in this equation is defined as
\begin{equation}
\mathcal{F}[\cdots]=\int d^2\bm{p}_T d^2\bm{k}_T\delta^2(\bm{p}_T+\bm{k}_T-\bm{q}_T)[\cdots],
\end{equation}
where $\bm{p}_T$, $\bm{k}_T$ are the transverse momenta of quarks in proton and antiproton,
and $\hat{\bm{h}}\equiv\frac{\bm{q}_T}{Q_T}$ is the direction of the
transverse momentum of $\gamma^*/Z$. The coefficients $K_1$, $K_3$ and $K_4$
are expressed as:
\begin{eqnarray}
K_1(\theta)&=&\frac{1}{4}(1+\cos^2\theta)[e_a^2+2g_V^le_ag_V^a\chi_1+c_1^lc_1^a\chi_2]\nonumber\\
&&+\frac{\cos\theta}{2}[2g_A^le_ag_A^a\chi_1+c_3^lc_3^a\chi_2],\\
K_3(\theta)&=&\frac{1}{4}\sin^2\theta[e_a^2+2g_V^le_ag_V^a\chi_1+c_1^lc_2^a\chi_2],\\
K_4(\theta)&=&\frac{1}{4}\sin^2\theta[2g_V^le_ag_A^a\chi_3],\label{K4}
\end{eqnarray}
where $e_a$ is the charge of quarks~(antiquarks), and $g_V$ and
$g_A$ are the vector and axial-vector coupling constants to the
$Z$-boson. We take their values in Ref.\cite{Nakamura:2010zzi}. The
$c_i$ is defined as:
\begin{equation}
\begin{split}
c_1^j=({g_V^j}^2+{g_A^j}^2)&,\quad c_2^j=({g_V^j}^2-{g_A^j}^2),\\
c_3^j=2g_V^jg_A^j&,
\end{split}
\end{equation}
where $j=l$ or $a$. The $Z$-boson propagator $\chi_i$ is given by:
\begin{eqnarray}
\chi_1&=&\frac{1}{\sin^2\theta_W}\frac{Q^2(Q^2-M_Z^2)}{(Q^2-M_Z^2)^2+\Gamma_Z^2M_Z^2},\\
\chi_2&=&\frac{1}{\sin^2\theta_W}\frac{Q^2}{Q^2-M_Z^2}\chi_1,\\
\chi_3&=&-\frac{\Gamma_Z M_Z}{Q^2-M_Z^2}\chi_1,\label{chi3}
\end{eqnarray}
where $\theta_W$ is the Weinberg angle. In Eq.(\ref{diffBoer}), we
assume that the TMD distribution functions for antiquarks~(quarks)
in the antiproton are the same as those for quarks~(antiquarks) in
proton, and the summation over the index $a$ are for different
flavors with $a=u$, $d$, $\bar{u}$, and $\bar{d}$.

In our calculation, we take the Boer-Mulders functions extracted
from $pD$ and $pp$ Drell-Yan data~\cite{Zhang:2008nu,Lu:2009ip}. The
parametrizations for $h_{1q}^{\perp}(x)$ is~\cite{Lu:2009ip}:
\begin{align}
h_{1q}^{\perp}(x)=H_qx^{c_q}(1-x)^bf_{1q}(x),
\end{align}
and the TMD part is parametrized with a Gaussian form:
\begin{eqnarray}
h_{1q}^{\perp}(x,\bm{p}_T^2)&=&h_{1q}^{\perp}(x)\frac{\exp(-\frac{\bm{p}_T^2}{p_{\mathrm{bm}}^2})}{\pi p_{\mathrm{bm}}^2},\\
f_{1q}(x,\bm{p}_T^2)&=&f_{1q}(x)\frac{\exp(-\frac{\bm{p}_T^2}{p_{\mathrm{un}}^2})}{\pi p_{\mathrm{un}}^2}.
\end{eqnarray}
This parametrization is based on the assumption that the $\cos2\phi$
asymmetry comes only from the Boer-Mulders effect in the region
$\bm{q}_T^2\ll Q^2$, and in this region the following relation hold:
\begin{eqnarray}
x_1=\frac{Q}{\sqrt{s}}e^y,\quad x_2=\frac{Q}{\sqrt{s}}e^{-y},
\end{eqnarray}
where $y$ is the rapidity of the $\gamma^*/Z$. We can also express
the cross section of the Drell-Yan process depending on $y$ and
$Q^2$ with an additional Jacobian determinant:
\begin{eqnarray}
\frac{d\sigma}{dydQ^2d^2\bm{q}_Td\Omega}=\frac{1}{s}\frac{d\sigma}{dx_1dx_2d^2\bm{q}_Td\Omega}.
\end{eqnarray}

From Eq.(\ref{diffBoer}), the azimuthal dependent terms are the
second and the third terms with $\cos2\phi$ and $\sin2\phi$ forms
respectively. However, the $\sin2\phi$ term is $\frac{1}{Q^2}$
suppressed, which can be found from (\ref{K4}) and (\ref{chi3}). As
shown in Ref.\cite{Lu:2011mz}, we can write the coefficient of
$\cos2\phi$ term in Eq.(\ref{diffcrosssectionW}) $W_{\Delta\Delta}$
into two parts, the perturbative QCD effect
$W_{\Delta\Delta}^{\mathrm{pQCD}}$ and the Boer-Mulders effect
$W_{\Delta\Delta}^{\mathrm{BM}}$. Then using an approximate Lam-Tung
relation $2W_{\Delta\Delta}^{\mathrm{pQCD}}-W_L\approx0$, one can
give the $\cos2\phi$ asymmetry caused by the Boer-Mulders effect:
\begin{eqnarray}
2\nu^{\mathrm{BM}}=\frac{4W_{\Delta\Delta}^{\mathrm{BM}}}{W_T+W_L}\approx2\nu+\lambda-1.
\end{eqnarray}
Comparing (\ref{diffcrosssectionW}) and (\ref{diffBoer}), and
neglecting the $W_L$ in the denominator because $W_L\ll W_T$ at low $\bm{q}_T$
region, we can get the following relation:
\begin{equation}\label{nu^BM}
\begin{split}
&\nu^{\mathrm{BM}}(\bm{q}_T,y,Q)=\\
&\frac{\sum_a\frac{1}{Q^2}K_3(\theta)\mathcal{F}
\left[(2\hat{\bm{h}}\cdot\bm{p}_T\hat{\bm{h}}\cdot\bm{k}_T-\bm{p}_T\cdot\bm{k}_T)
\frac{h_{1a}^{\perp}h_{1a}^{\perp}}{M^2}\right]}{\sum_a\frac{1}{Q^2}K_1(\theta)\mathcal{F}[f_{1a}f_{1a}]}.
\end{split}
\end{equation}

In the numerical calculation, we choose the values of parameters in
the Boer-Mulders function as those in
Ref.\cite{Lu:2009ip,Lu:2011cw}. There is still an unsettled factor
$\omega$ which might be flavor dependent in the parametrization,
because it will be canceled in the product of two Boer-Mulders
functions of quark and antiquark. It can range in the region
$0.48<\omega<2.1$, which is limited by the positivity
bounds~\cite{Bacchetta:1999kz,Zhang:2008nu,Lu:2009ip}. However, in
the $p\bar{p}$ Drell-Yan process, it has the product of two
Boer-Mulders functions of two quarks or two antiquarks which will
not cancel the factor $\omega$. The cross section has different
behavior with different values for $\omega$. Therefore, we can learn
additional information of the Boer-Mulders function from $p\bar{p}$
Drell-Yan processes. In this work, we choose three different values
for $\omega=0.5$, $\omega=1$ and $\omega=2$ to calculate
$\nu^{\mathrm{BM}}$ and show their different behavior.

In order to give $\nu^{\mathrm{BM}}$ with respect to a parameter
$y$, $Q$ or $q_T$, we should integrate for the other parameters of
the numerator and the denominator in Eq.(\ref{nu^BM}) respectively.
The integral over $q_T$ need to be cut off at $q_T=2~\textrm{GeV}$,
because intrinsic transverse momentum plays a significant role at
low $\bm{q}_T$ and the fitting for the parameters has excluded the
data with $q_T>2~\textrm{GeV}$.

\begin{figure}\label{bmq}
\includegraphics[width=0.47\textwidth]{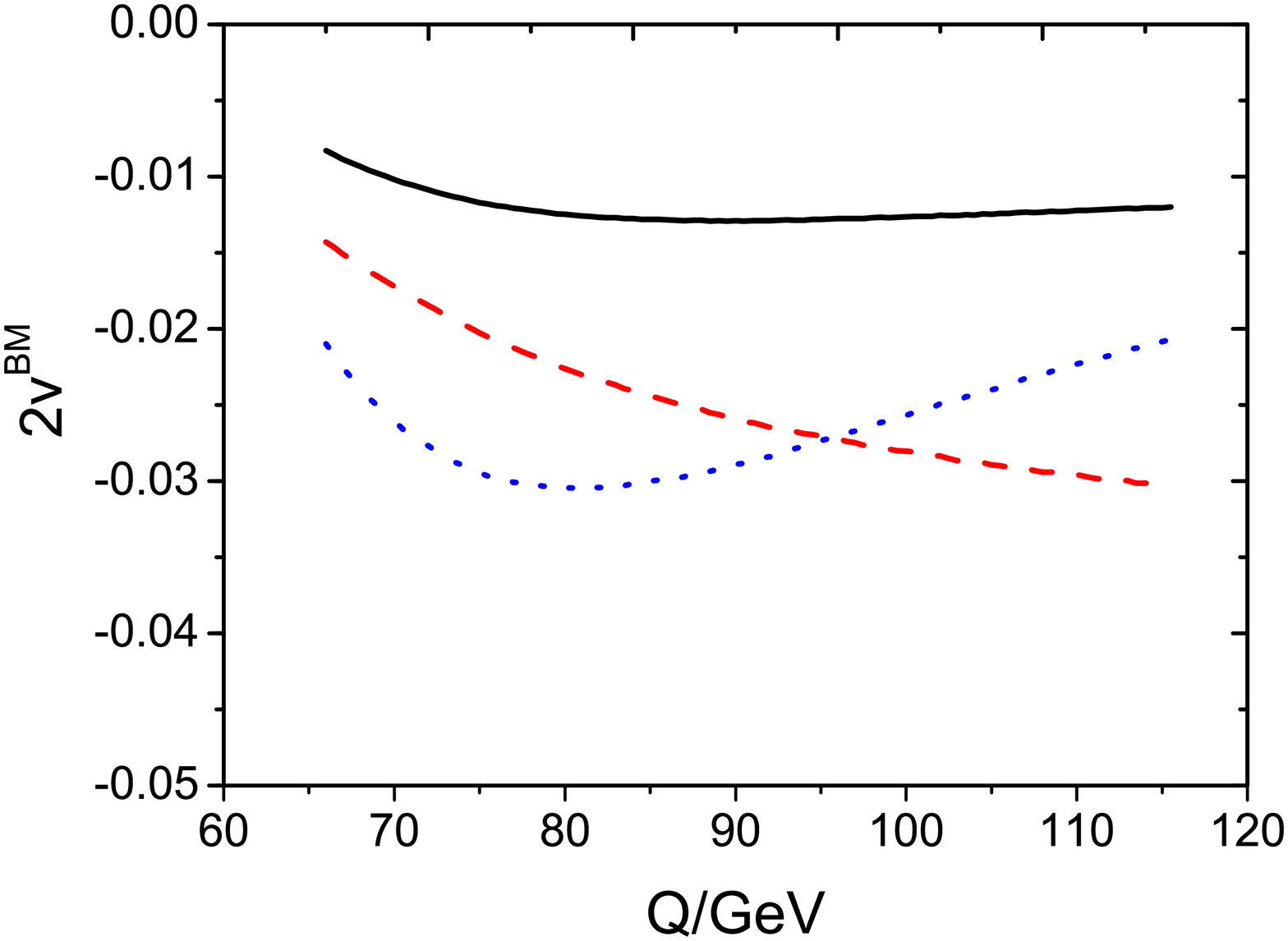}
\caption{The $Q$-dependent $\cos2\phi$ azimuthal asymmetry caused by
the Boer-Mulders effect in the unpolarized
$p+\bar{p}\rightarrow\gamma^*/Z+X\rightarrow l^+l^-+X$ process at
$Z$ mass region.  The solid, dashed, and dotted curves correspond to
$\omega=1$, $\omega=2$, and $\omega=0.5$ respectively.}
\end{figure}

\begin{figure}\label{bmy}
\includegraphics[width=0.47\textwidth]{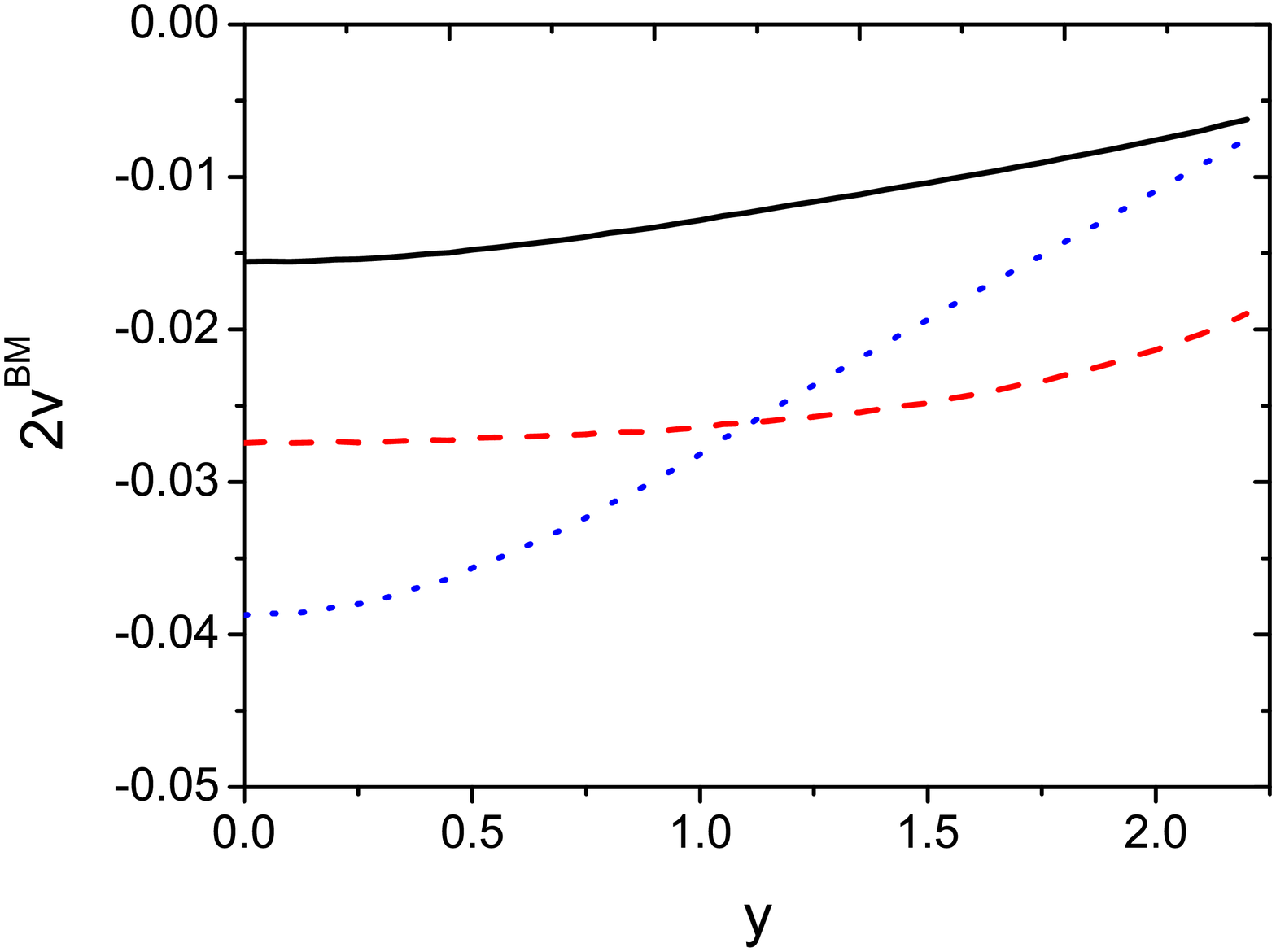}
\caption{The rapidity $y$-dependent $\cos2\phi$ azimuthal asymmetry
caused by the Boer-Mulders effect in the unpolarized
$p+\bar{p}\rightarrow\gamma^*/Z+X\rightarrow l^+l^-+X$ process at
$Z$ mass region.  The solid, dashed, and dotted curves correspond to
$\omega=1$, $\omega=2$, and $\omega=0.5$ respectively.}
\end{figure}

\begin{figure}\label{bmqt}
\includegraphics[width=0.47\textwidth]{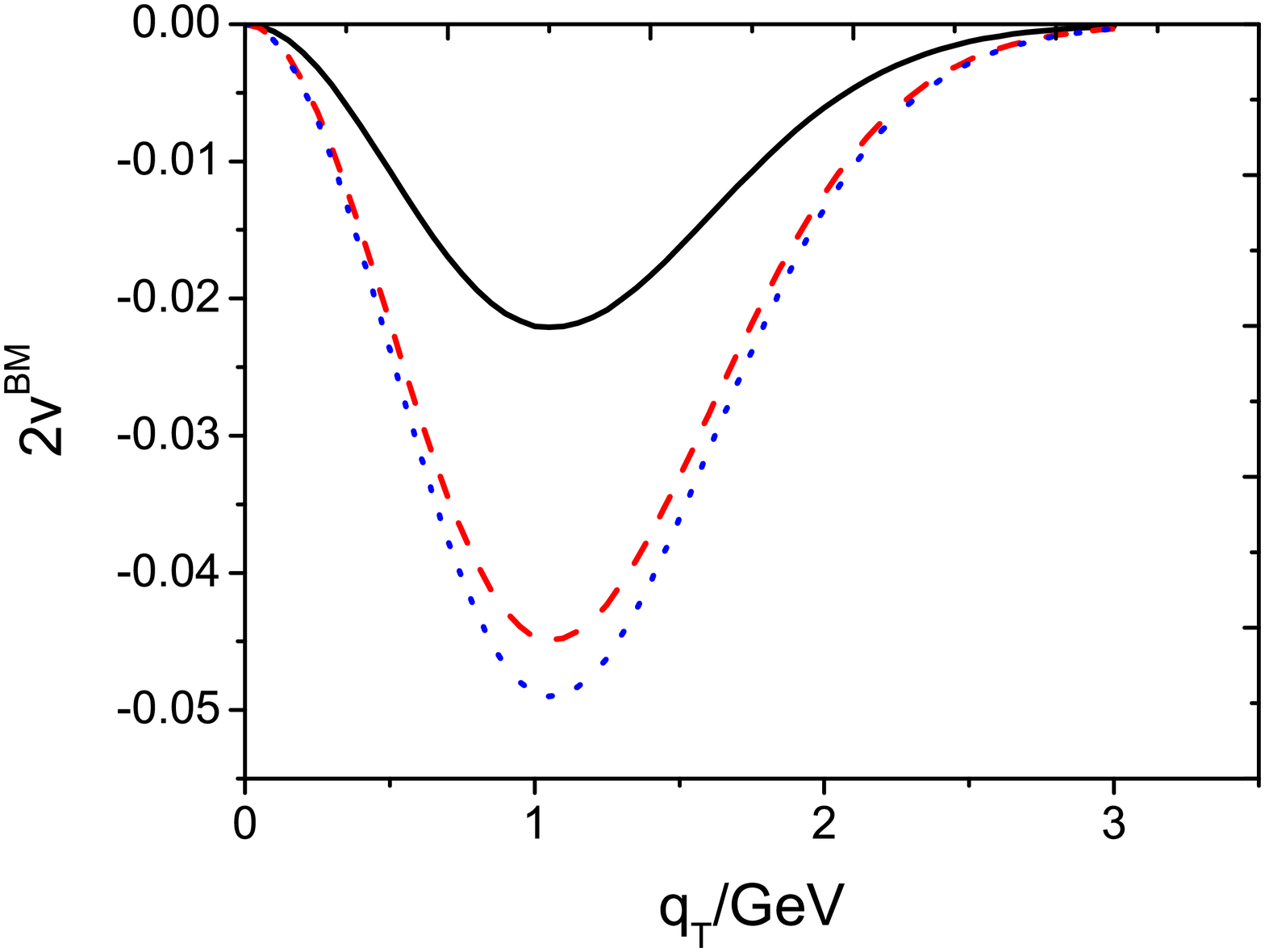}
\caption{The $q_T$-dependent $\cos2\phi$ azimuthal asymmetry caused
by the Boer-Mulders effect in the unpolarized
$p+\bar{p}\rightarrow\gamma^*/Z+X\rightarrow l^+l^-+X$ process at
$Z$ mass region. The solid, dashed, and dotted curves correspond to
$\omega=1$, $\omega=2$, and $\omega=0.5$ respectively.}
\end{figure}

Comparing (\ref{diffcrosssectionW}) with the angular distribution form taken by CDF~\cite{Aaltonen:2011nr}:
\begin{align}
\frac{d\sigma}{d\phi}\propto1+\beta_3\cos\phi+\beta_2\cos2\phi+\beta_7\sin\phi+\beta_5\sin2\phi,
\end{align}
$\nu^{\mathrm{BM}}$ will contribute to $\beta_2$ caused by the
Boer-Mulders effect at low $\bm{q}_T$.

In summary, we calculated the cos2$\phi$ azimuthal asymmetry in the
unpolarized $p\bar{p}$ Drell-Yan dilepton production processes in
the $Z$ mass region at CDF kinematic domain. It can be measured by
experimental detection of the Lam-Tung relation violation. It is
possible to study the spin structure of hadrons in unpolarized
collision processes around $Z$ mass region at Tevatron. In addition,
the $p\bar{p}$ processes can give more significant information of
the Boer-Mulders function than $pp$ processes. It can help us to
settle the factor $\omega$ in the parametrization, and the
prediction that the Boer-Mulders function have different signs in
SIDIS and Drell-Yan processes~\cite{Collins:2002kn} also awaits
experimental confirmation. Therefore the available data of Tevatron
are ideal to investigate the spin structure of nucleons via the
unpolarized $p\bar{p}$ process at the $Z$ pole. Besides, the
GSI-PANDA experiment~\cite{Lutz:2009ff} will run unpolarized
Drell-Yan processes with $p\bar{p}$ colliding at
$s=30~\mathrm{GeV}^2$, and PAX experiment~\cite{Barone:2005pu} may
preform unpolarized $p\bar{p}$ Drell-Yan process with the fixed
target mode at $s=45~\mathrm{GeV}^2$. They will provide us an
environment to study the Boer-Mulders effect at $J/\psi$ and
$\Upsilon$ peaks and to understand the structure of nucleons. All
these $p\bar{p}$ Drell-Yan experiments will give us significant
promotion in understanding the hadron structure and non-perturbative
QCD properties.

\begin{acknowledgments}
We are greatly indebted to Prof.~Liang Han for the stimulating
discussion about possible experimental analysis on spin physics at
Tevatron. This work is partially supported by National Natural
Science Foundation of China (Grants No.~11021092, No.~10975003,
No.~11035003, and No.~11120101004), by the Research Fund for the
Doctoral Program of Higher Education (China).
\end{acknowledgments}


\end{document}